\documentclass[conference]{IEEEtran}

\usepackage{cite}
\usepackage{amsmath,amsfonts,amssymb,bm}
\usepackage{graphicx}
\usepackage{booktabs}
\usepackage{multirow}
\usepackage{array}
\usepackage{url}

\graphicspath{{pics/}}
\newcommand{\E}{\mathbb{E}}
\newcommand{\R}{\mathbb{R}}
\newcommand{\C}{\mathbb{C}}
\newcommand{\cX}{\mathcal{X}}
\newcommand{\cL}{\mathcal{L}}

\begin{document}

\title{CoFi-CLM: A Coarse-to-Fine Channel Language Model 
for Finite-Bit CSI Feedback}

\author{
\IEEEauthorblockN{Chao Zhang, Cheng Luo\textsuperscript{*}, Luping Xiang, and Kun Yang}
\IEEEauthorblockA{State Key Laboratory of Novel Software Technology, Nanjing University, Nanjing 210008, China,\\
and School of Intelligent Software and Engineering, Nanjing University (Suzhou Campus), Suzhou, China.\\
\textsuperscript{*}Corresponding Author, Email: chengluo@nju.edu.cn}
}

\maketitle

\begin{abstract}
In frequency-division duplex massive multiple-input multiple-output (MIMO) systems,
the user equipment (UE) must convey high-dimensional downlink channel state
information (CSI) under a stringent finite-bit feedback budget. Deep learning (DL) has emerged as a powerful tool for CSI compression due to its ability to capture complex channel correlations and learn compact CSI representations. However, recovering fine-scale channel structure from limited feedback remains challenging. To address this challenge, we propose the Coarse-to-Fine Channel Language Model (CoFi-CLM), a large AI model deployed at the base station (BS) that explicitly learns dependencies between coarse and fine CSI representations. Specifically, the UE reports coarse-token indices from a learned codebook, while CoFi-CLM predicts distributions over unreported fine-token indices conditioned on this feedback in a single Transformer forward pass. A dual-path decoder fuses the direct coarse
reconstruction with the generated fine-scale reconstruction. Concentrating
fine-token prediction and fusion at the BS allows the CLM capacity to scale
without increasing the computational or storage cost at the UE. Across 64 to 160 feedback bits, CoFi-CLM consistently outperforms the compared methods on both seen and unseen scenarios. At 128 bits, it improves the normalized mean square error (NMSE) over the recent large AI model baseline by approximately 1 dB on both sets. The performance gap between seen and unseen scenarios further supports its generalization under the considered channel model. The
source code is publicly available at
\url{https://github.com/doovvv/CoFi-CLM}.
\end{abstract}


\section{Introduction}
In frequency-division duplex (FDD) massive multiple-input multiple-output
(MIMO) systems, the base station (BS) relies on downlink channel state
information (CSI) reported by the user equipment (UE) for precoding because
direct uplink--downlink reciprocity is unavailable. The feedback overhead of
raw CSI grows rapidly with the numbers of antennas and subcarriers
\cite{mashhadi2020csi}.

Conventional CSI feedback mainly adopts codebook-based quantization or
compressive sensing (CS). The former quantizes channel directions or precoders
with a predefined finite codebook and feeds back the index of the selected entry
\cite{love2008limitedfeedback}. Maintaining adequate quantization resolution as
the number of BS antennas grows requires a larger codebook, increasing storage and
search complexity \cite{mashhadi2020csi}. CS-based methods instead exploit the approximate sparsity of
massive-MIMO channels in the angular--delay domain and common support across
subcarriers \cite{gao2015spatial}. Their reconstruction performance depends on
the accuracy of the assumed sparse model, while iterative recovery incurs
additional computational cost \cite{mashhadi2020csi}.

Data-driven CSI feedback instead learns the compression mapping from channel
samples. CsiNet introduced an autoencoder that maps the angle--delay-domain CSI
matrix to a low-dimensional latent vector and reconstructs it at the BS
\cite{wen2018csinet}. Subsequent work improves learned CSI compression and 
reconstruction through deeper neural architectures and attention mechanisms \cite{guo2020csinetplus,cui2022transnet}.

Beyond direct reconstruction from continuous latent vectors, sequence-based
and generative CSI representations have also been explored. Sequence-based
methods organize CSI features as sequences for Transformer processing
\cite{cui2022transnet}. Masked-token Transformers recover omitted CSI features
using learnable mask tokens \cite{zhao2023maskedtoken}. Diffusion-based methods
exploit generative priors for CSI reconstruction
\cite{kim2025diffusion,cheng2026csicogen}. However, three issues remain when
these methods are applied to coarse-to-fine finite-bit feedback. First, the
sequence elements are not selected from a finite learned vocabulary of
recurring CSI patterns. Second, masked-token Transformers operate within a
single
representation scale and do not explicitly model the mapping from limited
coarse tokens to fine-scale CSI structure. Third, diffusion-based
reconstruction typically requires multiple denoising steps at inference.

To address these limitations, we propose the Coarse-to-Fine Channel Language Model (CoFi-CLM), a BS-side channel language model (CLM) for conditional fine-token prediction. It learns cross-scale dependencies from coarse feedback to fine-resolution CSI and combines the generated reconstruction with direct coarse decoding.
The main contributions are summarized as follows.
\begin{itemize}
    \item We formulate finite-bit CSI recovery as predicting fine-token
    distributions conditioned on a finite-bit
    coarse-token context. Two CSI tokenizers learned at different spatial
    resolutions provide separate coarse and fine codebooks. With these
    codebooks, the conditional prediction task explicitly models cross-scale
    dependencies between the two discrete CSI representations.
    \item We develop an asymmetric CoFi-CLM architecture in which the UE
    reports coarse-token indices and the BS predicts all fine-token distributions
    in parallel within one Transformer forward pass, avoiding autoregressive
    decoding and iterative diffusion sampling. The generated reconstruction is
    then fused with direct coarse decoding to exploit their complementary
    information. Concentrating prediction and fusion at the BS allows model
    capacity to scale without increasing UE computation or storage.
    \item We evaluate CoFi-CLM at six actual feedback budgets from 64 to
    160 bits on seen and unseen scenarios. The large CLM outperforms the
    strongest baseline at every evaluated rate by 0.490--1.245 dB. The
    parameter-matched comparison indicates that the gain cannot be explained
    by model size alone. The reconstruction-path comparison shows that the
    generated fine-scale information complements direct coarse reconstruction.
    At 128 bits, scaling only the BS-side CLM yields a further
    0.195--0.203-dB improvement, and the complexity analysis confirms that this
    scaling leaves UE computation and storage unchanged.
\end{itemize}

% \emph{Notation:} Bold lowercase and uppercase letters denote vectors and
% matrices or multidimensional arrays, respectively. $\C$ and $\R$ denote the
% complex and real fields. $(\cdot)^H$, $\|\cdot\|_{\rm F}$, and $\E[\cdot]$
% denote the Hermitian transpose, Frobenius norm, and expectation, respectively.

\section{System Model and Problem Formulation}
\subsection{FDD CSI Feedback Model}
Consider an FDD downlink in which a BS equipped with $N_t$ transmit antennas
serves a single-antenna UE over $N_f$ orthogonal frequency-division
multiplexing (OFDM) subcarriers. On subcarrier
$n\in\{0,\ldots,N_f-1\}$, the received complex-baseband symbol is
\begin{equation}
\label{eq:received_signal}
    y_n=\mathbf h_n^H\mathbf v_n s_n+w_n,
\end{equation}
where $\mathbf h_n\in\C^{N_t}$ is the downlink channel vector,
$\mathbf v_n\in\C^{N_t}$ is the precoder, $s_n\in\C$ is the transmitted
symbol, and $w_n\in\C$ is additive noise. Downlink precoding requires the BS
to obtain $\mathbf h_n$ for all subcarriers.

Stacking the channel vectors by rows yields the spatial--frequency CSI matrix
\begin{equation}
\label{eq:spatial_frequency_csi}
    \mathbf H=[\mathbf h_0,\ldots,\mathbf h_{N_f-1}]^H
    \in\C^{N_f\times N_t}.
\end{equation}
Because channel energy is typically more concentrated in the angle--delay
domain \cite{wen2018csinet}, a two-dimensional discrete Fourier transform is applied along the
subcarrier and antenna dimensions, after which the effective region is
retained. Let $\widetilde{\mathbf H}\in\C^{N_d\times N_a}$ denote the truncated
angle--delay-domain CSI, where $N_d$ and $N_a$ are the retained delay and
angular dimensions.

To form the real-valued neural-network input, each complex angle--delay-domain
CSI matrix $\widetilde{\mathbf H}$ is converted into a two-channel tensor
$\mathbf X$. The first and second channels contain the normalized real and
imaginary parts, respectively, while the last two dimensions correspond to
delay and angle. Let $\alpha>0$ denote the maximum absolute value among the
retained real and imaginary CSI coefficients in the data used for
preprocessing. Dividing by $\alpha$ scales both parts to $[-1,1]$, after which
they are shifted and rescaled to $[0,1]$ as
\begin{equation}
\label{eq:csi_preprocessing}
\begin{split}
    &\mathbf X_{1,:,:}=\frac{1}{2}
    \left(\frac{\operatorname{Re}(\widetilde{\mathbf H})}{\alpha}+1\right),\\
    &\mathbf X_{2,:,:}=\frac{1}{2}
    \left(\frac{\operatorname{Im}(\widetilde{\mathbf H})}{\alpha}+1\right).
\end{split}
\end{equation}
Thus, $\mathbf X\in[0,1]^{2\times N_d\times N_a}$ is the real-valued CSI
representation supplied to the neural networks. We set $N_t=N_a=32$ and retain
$N_d=32$ delay coefficients, giving an input size of $2\times32\times32$.

\subsection{Finite-Bit CSI Reconstruction}
Let $B\in\mathbb N$ be the maximum number of feedback bits per CSI sample. A UE
encoder $f_{\rm UE}$ maps $\mathbf X$ to a binary feedback vector $\mathbf b$,
and a BS decoder $g_{\rm BS}$ produces $\widehat{\mathbf X}$. This finite-bit
interface is expressed as
\begin{equation}
\label{eq:finite_bit_interface}
    \mathbf b=f_{\rm UE}(\mathbf X)\in\{0,1\}^{B_{\rm act}},
    \quad B_{\rm act}\leq B,\quad
    \widehat{\mathbf X}=g_{\rm BS}(\mathbf b),
\end{equation}
where $\widehat{\mathbf X}\in\R^{2\times N_d\times N_a}$ and $B_{\rm act}$ is
the actual feedback length. An error-free feedback link is assumed to isolate the effect of the
digital bit budget.

The finite-bit CSI feedback problem jointly designs $f_{\rm UE}$ and
$g_{\rm BS}$ to minimize the CSI reconstruction error after compression under
the feedback budget. For the channel-sample distribution $p_{\mathbf X}$, this
problem is formulated as
\begin{equation}
\label{eq:finite_bit_problem}
\begin{aligned}
    \min_{f_{\rm UE},g_{\rm BS}}\quad&
    \E_{\mathbf X\sim p_{\mathbf X}}
    \left[\left\|\mathbf X-
    g_{\rm BS}\!\left(f_{\rm UE}(\mathbf X)\right)\right\|_{\rm F}^{2}\right]\\
    \text{s.t.}\quad& B_{\rm act}\leq B .
\end{aligned}
\end{equation}
A lower value of $B$ reduces the CSI feedback overhead but also limits the
information available for BS-side reconstruction. This work evaluates
reconstruction accuracy under a range of stringent feedback-bit budgets.

\subsection{Evaluation Metric}
Let $\cX=\{\mathbf X^{(q)}\}_{q=1}^{N_{\rm test}}$ denote the test set,
where $N_{\rm test}$ is the number of test samples. Since
Eq. \eqref{eq:csi_preprocessing} maps a zero complex channel coefficient to
$0.5$, we first center the target and reconstructed samples elementwise as
$\bar{\mathbf X}^{(q)}=\mathbf X^{(q)}-0.5$ and
$\widehat{\bar{\mathbf X}}^{(q)}=\widehat{\mathbf X}^{(q)}-0.5$,
respectively. Using these centered CSI samples, we measure reconstruction
accuracy by the global power-ratio normalized mean square error (NMSE)
\begin{equation}
\label{eq:nmse}
    {\rm NMSE}_{\rm dB}=10\log_{10}
    \frac{\sum_{q=1}^{N_{\rm test}}
    \|\bar{\mathbf X}^{(q)}-\widehat{\bar{\mathbf X}}^{(q)}\|_{\rm F}^{2}}
    {\sum_{q=1}^{N_{\rm test}}\|\bar{\mathbf X}^{(q)}\|_{\rm F}^{2}}.
\end{equation}
A lower NMSE indicates more accurate CSI reconstruction.

\section{Large AI Channel Language Model}
\begin{figure*}[t]
    \setlength{\abovecaptionskip}{2pt}
    \setlength{\belowcaptionskip}{0pt}
    \centering
    \includegraphics[width=0.98\textwidth]{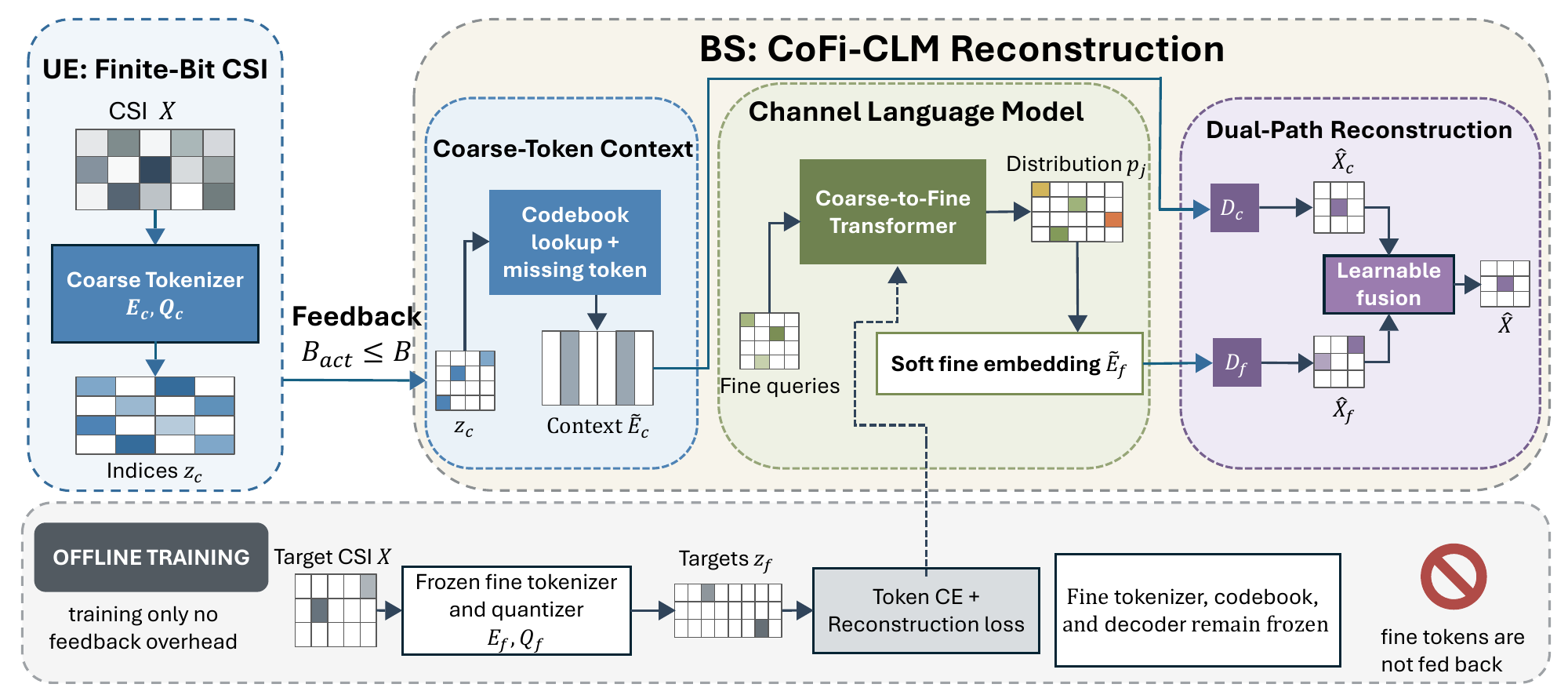}
    \caption{CoFi-CLM architecture and asymmetric deployment. The upper part shows online CSI feedback and reconstruction, while the dashed lower part provides fine-token targets for offline supervision.}
    \label{fig:architecture}
\end{figure*}

\subsection{Overview and Asymmetric Deployment}
Fig.~\ref{fig:architecture} shows the complete processing pipeline of the
proposed BS-side large AI channel language model. The upper part depicts the
online feedback process, which comprises coarse-token extraction, finite-bit
transmission, BS-side fine-token generation, and dual-path reconstruction. The
UE runs only a lightweight coarse
tokenizer, which converts the current CSI into a $4\times4$ grid of coarse-token
indices and transmits a budget-dependent subset. Using the received indices,
the BS constructs a $4\times4$ coarse-token context and decodes it through the
direct coarse reconstruction path. In parallel, this context conditions the
CLM, which predicts an $8\times8$ grid of
categorical distributions over the fine codebook in one forward pass. A
learnable fusion module operates on the two reconstruction features to produce
the final CSI estimate.

The fine tokenizer is pretrained separately and kept frozen during CLM training. Its encoder and quantizer, shown in the dashed lower part of Fig.~\ref{fig:architecture}, map the ground-truth CSI to fine-token indices on a higher-resolution spatial grid to provide supervision. During online operation, the CLM uses the reported coarse-token indices of the current sample to construct its context and exploits a learned channel prior to predict fine-token distributions. The frozen fine decoder maps the resulting soft fine embeddings to a fine-scale CSI reconstruction. The main generation workload therefore remains at the BS, while the UE performs only coarse encoding and codebook matching.
\subsection{Coarse and Fine Tokenizers}
The coarse and fine tokenizers represent CSI at different spatial resolutions
using separate learned codebooks. Each \emph{codebook} contains a finite set of
learned vectors, referred to as \emph{tokens}. A \emph{token index} is the integer
identifying a token within its codebook. Each tokenizer assigns each spatial
latent vector to a token and outputs its index. Throughout this paper, token
feedback refers to transmitting the corresponding indices, and token prediction
refers to predicting categorical distributions over those indices. In this
section, $i$ and $j$ denote spatial indices, whereas a token index identifies
an entry in the corresponding codebook.

The coarse tokenizer comprises an encoder $E_c$, a vector quantizer $Q_c$, and
a decoder $D_c$. For the $32\times32$ spatial CSI grid, it uses an effective
$8\times8$ patch size and produces a $4\times4$ grid of $L_c=16$ coarse
tokens. The encoder $E_c$ maps $\mathbf X$ to 16 latent vectors of dimension
64. For the $i$th latent
$\mathbf e_{c,i}$, $Q_c$ selects the nearest entry from a codebook containing
$K_c$ vectors
\begin{equation}
\label{eq:vq}
    z_{c,i}=\arg\min_{k\in\{1,\ldots,K_c\}}
    \|\mathbf e_{c,i}-\mathbf c_{c,k}\|_2^2 ,
\end{equation}
where $\mathbf c_{c,k}\in\R^{64}$ is the $k$th coarse token. The integer
$z_{c,i}$ is the coarse-token index at spatial index $i$, and
$\mathbf c_{c,z_{c,i}}$ is the corresponding quantized latent vector. The
ordered vector $\mathbf z_c=(z_{c,1},\ldots,z_{c,L_c})$ is the coarse-token index
sequence. The coarse codebook is
updated by exponential moving averages and trained jointly with the remaining
modules. Because the UE and BS share this codebook, transmitting an integer
index is sufficient for the BS to recover its 64-dimensional token.

The fine tokenizer represents CSI at a higher spatial resolution than the
coarse tokenizer. It is pretrained independently and comprises an encoder $E_f$,
a vector quantizer $Q_f$, and a decoder $D_f$. It
uses an effective $4\times4$ patch size and produces an $8\times8$ grid of
$L_f=64$ fine tokens. First, the encoder transforms $\mathbf X$ into the
fine-scale latent matrix
\begin{equation}
\label{eq:fine_encoding}
    \mathbf E_f=E_f(\mathbf X)
    =[\mathbf e_{f,1},\ldots,\mathbf e_{f,L_f}]^{\mathsf T}
    \in\R^{L_f\times64}.
\end{equation}
Each latent vector $\mathbf e_{f,j}$ corresponds to spatial index $j$ in
the $8\times8$ fine-token grid. Let the fine codebook contain $K_f=256$
tokens $\mathbf c_{f,k}\in\R^{64}$. The vector quantizer then assigns each
latent vector to its nearest fine token. The corresponding index is
\begin{equation}
\label{eq:fine_quantization}
    z_{f,j}=\arg\min_{k\in\{1,\ldots,K_f\}}
    \|\mathbf e_{f,j}-\mathbf c_{f,k}\|_2^2,
    \quad j=1,\ldots,L_f.
\end{equation}
The ordered vector
$\mathbf z_f=(z_{f,1},\ldots,z_{f,L_f})\in\{1,\ldots,K_f\}^{L_f}$ is the
fine-token index sequence. Direct feedback of the complete sequence would
require $L_f\lceil\log_2K_f\rceil=512$ bits, which exceeds the considered
64--160-bit budgets. CoFi-CLM therefore uses $\mathbf z_f$ as a BS-side
prediction target rather than including it in the feedback payload. Retrieving the token for each index gives a quantized latent matrix,
which the decoder $D_f$ maps back to the CSI domain as
\begin{equation}
\label{eq:fine_decoding}
    \widehat{\mathbf X}_{\rm oracle}
    =D_f\!\left([\mathbf c_{f,z_{f,1}},\ldots,
    \mathbf c_{f,z_{f,L_f}}]^{\mathsf T}\right).
\end{equation}
This oracle reconstruction uses all fine-token indices and is therefore not
subject to the feedback-bit constraint. After pretraining, $E_f$, $Q_f$,
$D_f$, and the fine codebook are frozen. The two codebooks are learned from
patterns at different spatial scales and do not share entries or index
semantics.

\subsection{Finite-Bit Coarse-Token Context}
\label{sec:budget}
Let $\mathcal S_B\subseteq\{1,\ldots,L_c\}$ denote the set of spatial indices
selected for feedback under budget $B$, with $N_s=|\mathcal S_B|$. For each
$i\in\mathcal S_B$, transmitting the coarse-token index $z_{c,i}$ requires
$b_c=\lceil\log_2K_c\rceil$ bits, so the actual overhead is
\begin{equation}
\label{eq:bitbudget}
    B_{\rm act}=N_s b_c\leq B .
\end{equation}
When the budget is insufficient for all $L_c$ coarse-token indices, nested
selection sets $\mathcal S_B$ are constructed from the importance of each
spatial index measured on the training data.

At the BS, coarse-token entries whose spatial indices are not in $\mathcal S_B$
are filled with a shared learnable missing embedding $\mathbf m_c\in\R^{64}$:
\begin{equation}
\label{eq:missing_embedding}
\widetilde{\mathbf e}_{c,i}=
\begin{cases}
\mathbf c_{c,z_{c,i}}, & i\in\mathcal S_B,\\
\mathbf m_c, & i\notin\mathcal S_B .
\end{cases}
\end{equation}
The vector $\mathbf m_c$ represents an unreported coarse-token entry and
contains no sample-specific channel value. Assigning either a retrieved token
or the missing embedding to each of the $L_c$ spatial positions yields a
fixed-length context $\widetilde{\mathbf E}_c\in\R^{L_c\times64}$.
The BS fills the missing positions locally, without increasing the feedback
overhead in Eq. \eqref{eq:bitbudget}.

\subsection{BS-Side Conditional Fine-Token Prediction}
The core task of the CLM is to predict fine-token distributions for the current
CSI sample from its finite-bit coarse-token context. During training, the frozen
fine tokenizer extracts the target indices from the ground-truth CSI.
The CLM with parameters $\theta$ projects the coarse context to $d$ dimensions
and adds two-dimensional positional embeddings to form $L_c$ coarse memory
embeddings. In parallel, the projected coarse feature map is
bilinearly interpolated to an $8\times8$ grid and added to $L_f=64$ learnable
fine queries and their positional embeddings. Each Transformer decoder layer
applies noncausal self-attention among the fine queries and cross-attention to
the entire coarse memory.

The final representation at spatial index $j$ produces logits
$\boldsymbol\ell_j\in\R^{K_f}$. For $k\in\{1,\ldots,K_f\}$, the conditional
probability and the corresponding soft fine embedding are
\begin{equation}
\label{eq:conditional_token}
\begin{aligned}
    &p_{j,k}\triangleq
    \Pr_{\theta}\!\left(z_{f,j}=k\mid\widetilde{\mathbf E}_c\right)
    =\frac{\exp(\ell_{j,k})}
    {\sum_{r=1}^{K_f}\exp(\ell_{j,r})},\\
    &\widetilde{\mathbf e}_{f,j}
    =\sum_{k=1}^{K_f}p_{j,k}\mathbf c_{f,k}.
\end{aligned}
\end{equation}
Stacking the fine embeddings in spatial-index order gives
$\widetilde{\mathbf E}_f\in\R^{L_f\times64}$. All $L_f$ fine-token distributions
are predicted in one forward pass, requiring neither autoregressive decoding
nor iterative diffusion sampling.

\subsection{Dual-Path Reconstruction and Training Objective}
Under finite-bit feedback, the two reconstruction paths have complementary
roles. The coarse path reconstructs CSI directly from the received $4\times4$
token context. The fine path instead uses the $8\times8$ token representation
predicted at the BS, which represents CSI at a higher spatial resolution but
may contain token-prediction errors. The fusion module therefore combines both
estimates rather than replacing the directly decoded coarse result with the
generated one. The direct coarse and generated fine reconstructions are
\begin{equation}
\label{eq:coarse}
    \widehat{\mathbf X}_c=D_c(\widetilde{\mathbf E}_c),
\end{equation}
and
\begin{equation}
\label{eq:fine}
    \widehat{\mathbf X}_f=D_f(\widetilde{\mathbf E}_f),
\end{equation}
respectively. Both $\widehat{\mathbf X}_c$ and $\widehat{\mathbf X}_f$ belong to
$\R^{2\times N_d\times N_a}$, where the first dimension contains the real and
imaginary CSI channels. Concatenating $\widehat{\mathbf X}_c$,
$\widehat{\mathbf X}_f$, and $\widehat{\mathbf X}_f-\widehat{\mathbf X}_c$
along this first dimension yields $\mathbf s\in\R^{6\times N_d\times N_a}$.
A spatial gate $\mathbf G_\psi(\mathbf s)\in\R^{2\times N_d\times N_a}$ and a
residual corrector $\mathbf R_\psi(\mathbf s)\in\R^{2\times N_d\times N_a}$
produce
\begin{equation}
\label{eq:fusion}
    \widehat{\mathbf X}
    =\widehat{\mathbf X}_c
    +\mathbf G_\psi(\mathbf s)\odot
    (\widehat{\mathbf X}_f-\widehat{\mathbf X}_c)
    +\mathbf R_\psi(\mathbf s).
\end{equation}

Training jointly optimizes CSI reconstruction and fine-token prediction.
For the latter, the frozen fine tokenizer provides the target index $z_{f,j}$
at each fine spatial position $j$. The CLM assigns probability $p_{j,z_{f,j}}$
to this target. To encourage higher probability for the target token, we
minimize the average cross-entropy loss over all $L_f$ fine positions,
\begin{equation}
\label{eq:tokenloss}
    \cL_{\rm token}=-\frac{1}{L_f}\sum_{j=1}^{L_f}
    \log p_{j,z_{f,j}} .
\end{equation}
Let $\cL_{\rm rec}$, $\cL_{\rm coarse}$, and $\cL_{\rm fine}$ denote
reconstruction losses for the fused, coarse, and generated fine outputs,
respectively. Each combines MSE with log-domain NMSE. The branch-level losses
encourage both paths to remain individually informative before fusion.
$\cL_{\rm VQ}$ is the coarse-tokenizer commitment loss. Codebook-diversity and
latent-variance terms form
$\cL_{\rm reg}=\lambda_{\rm div}\cL_{\rm div}
+\lambda_{\rm var}\cL_{\rm var}$. The complete objective is
\begin{equation}
\label{eq:loss}
\begin{split}
    \cL={}&\lambda_r\cL_{\rm rec}
    +\lambda_c\cL_{\rm coarse}
    +\lambda_f\cL_{\rm fine}
    +\lambda_{\rm CE}\cL_{\rm token}\\
    &+\lambda_{\rm VQ}\cL_{\rm VQ}+\cL_{\rm reg}.
\end{split}
\end{equation}
The reconstruction terms optimize CSI fidelity, while $\cL_{\rm token}$ learns
the conditional categorical distribution.

\section{Experimental Results}
\subsection{Dataset and Evaluation Protocol}
Channels are generated using QuaDRiGa \cite{jaeckel2014quadriga} under the
3GPP TR~38.901 urban microcell non-line-of-sight model \cite{3gpp38901}. The
carrier frequency is 2 GHz, the bandwidth is 10 MHz, and the BS uses 32
antennas over 64 subcarriers. The dataset contains 3,200 scenarios with 400
samples per scenario, for a total of 1.28 million samples. From 3,000 seen
scenarios, 945,000 samples are used for training, 105,000 for validation, and
150,000 for testing. The remaining 200 scenarios are reserved for unseen-scene
evaluation.

For each unseen scenario, the last 50 samples form a common query set, giving
10,000 query samples. These queries are disjoint from the other 350 samples in
the same scenario and from all seen-scenario data. All methods assume an
error-free feedback link and are evaluated using Eq. \eqref{eq:nmse}.

\subsection{Implementation Details and Baselines}
The fine tokenizer is pretrained to reconstruct absolute CSI and remains
frozen during CLM training. For coarse feedback, the
64-, 96-, and 112-bit settings use $K_c=256$ and transmit 8, 12, and 14
coarse-token indices, respectively. The 128-bit setting transmits all 16 indices
with $K_c=256$, while the 144- and 160-bit settings transmit all 16 indices with
$K_c=512$ and $K_c=1024$, respectively.

The large CLM has a 384-dimensional, 12-layer, 12-head Transformer with an
inner feed-forward dimension of 1,536. We set $\lambda_r=1$,
$\lambda_c=0.25$, $\lambda_f=0.1$, $\lambda_{\rm CE}=0.02$,
$\lambda_{\rm VQ}=0.25$, $\lambda_{\rm div}=0.05$, and
$\lambda_{\rm var}=0.05$. AdamW is used for 300,000 steps with batch size 256. The learning
rate increases linearly to $8\times10^{-5}$ over 1,500 warm-up steps and then
follows cosine decay to $4\times10^{-6}$. All CLM runs use seed 48, mixed-precision training, and gradient checkpointing.

We compare six models. The first two are a tiny AI model with one Transformer
layer retained only in the decoder and a small AI model with one Transformer
layer in both the encoder and decoder. The third and fourth follow the recent
large AI model (LAM) CSI-feedback framework in \cite{guo2025promptlam}, which reconstructs
CSI at the BS from a continuous latent vector with each element quantized to 4 bits.
LAM w/o Prompt uses only the quantized latent vector, whereas Prompt-LAM
additionally conditions the BS decoder on a scene-average-magnitude
prompt computed from support CSI. The fifth and sixth are a five-layer small
CLM and the proposed 12-layer large CLM, respectively. For each unseen
scenario, the first 350 samples form the support set, which is disjoint from
the query set.
\begin{figure*}[t]
    \setlength{\abovecaptionskip}{2pt}
    \setlength{\belowcaptionskip}{0pt}
    \centering
    \includegraphics[width=0.92\textwidth]{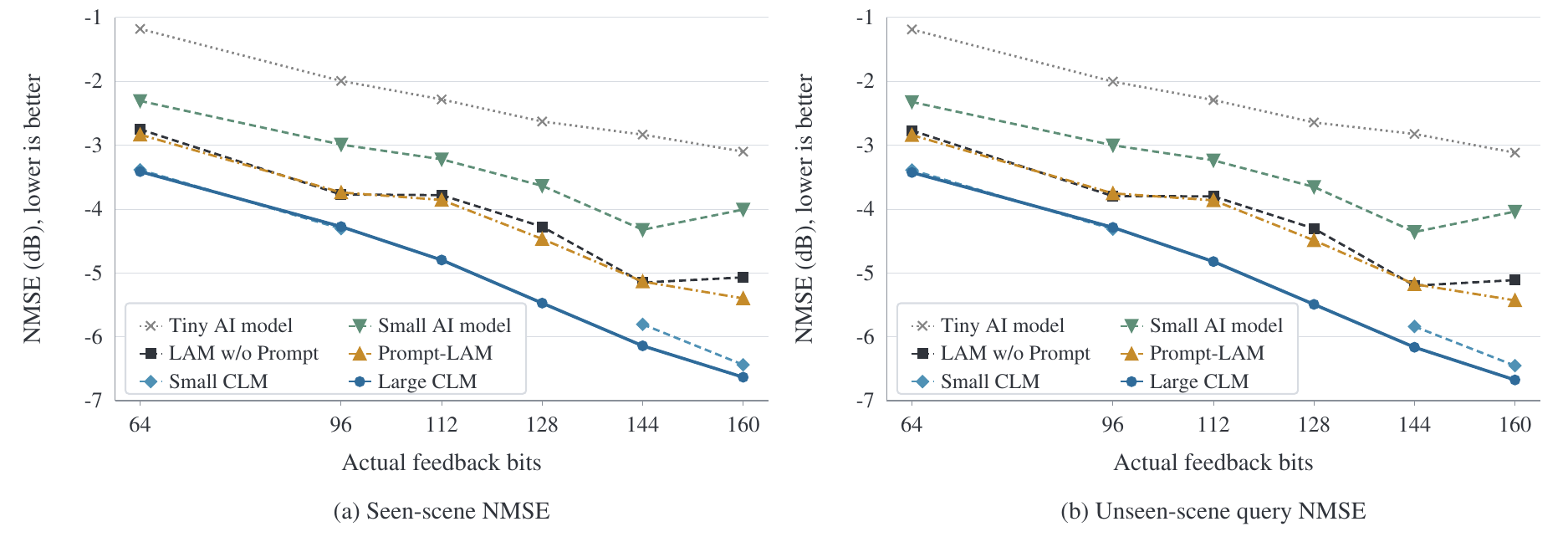}
    \caption{CSI reconstruction performance from 64 to 160 feedback bits.
    Panel (a) uses 150,000 seen-scenario test samples. In panel (b), all
    methods use the same 10,000 unseen-scenario query samples.}
    \label{fig:rate}
\end{figure*}

\subsection{Performance at Different Feedback Rates}
Fig.~\ref{fig:rate} compares NMSE across six feedback budgets. The large CLM
outperforms the external baselines on seen and unseen scenarios at all six
rates. The small CLM also outperforms the external baselines at all six
feedback budgets. At 128 bits, the large CLM achieves $-5.4708$
dB on seen scenarios and $-5.4907$ dB on unseen queries, improving upon
Prompt-LAM by 1.004 dB and 1.002 dB, respectively.

Beyond these reconstruction gains, the large CLM exhibits a maximum absolute
seen--unseen NMSE difference of only 0.043 dB across all six rates. This small
gap supports generalization to new scatterer realizations under the considered
channel model. These unseen-scenario results are obtained without
scenario-specific support data or online adaptation, since fine-token
prediction depends only on the current sample's coarse context and the model
learned offline.

\begin{table}[t]
    \centering
    \caption{Parameter-Matched Comparison and BS-Side Scaling at 128 Bits}
    \label{tab:param}
    \setlength{\tabcolsep}{3.5pt}
    \begin{tabular}{lrrr}
        \toprule
        \multirow{2}{*}{Method} & \multirow{2}{*}{Parameters}
        & \multicolumn{2}{c}{NMSE (dB)}\\
        \cmidrule(lr){3-4}
        & & Seen & Unseen-Q\\
        \midrule
        Prompt-LAM & 5.676M & $-4.4665$ & $-4.4888$\\
        Small CLM & 5.678M & $-5.2757$ & $-5.2872$\\
        Large CLM & 31.448M & $\mathbf{-5.4708}$ & $\mathbf{-5.4907}$\\
        \bottomrule
    \end{tabular}
\end{table}

\subsection{Parameter-Matched Comparison and BS-Side Scaling}
Table~\ref{tab:param} lists the 128-bit models. The small CLM and Prompt-LAM
have nearly identical parameter counts of 5.678M and 5.676M, respectively.
The small CLM improves the seen and unseen-query NMSE over Prompt-LAM by
0.809 dB and 0.798 dB, respectively. This parameter-matched comparison
indicates that the gain cannot be explained by model size alone. Scaling only
the BS-side CLM provides further improvements of 0.195 dB and 0.203 dB while
leaving the UE architecture unchanged. This additional gain characterizes a
performance--complexity tradeoff confined to the BS.

\begin{figure}[t]
    \setlength{\abovecaptionskip}{2pt}
    \setlength{\belowcaptionskip}{0pt}
    \centering
    \includegraphics[width=0.98\columnwidth]{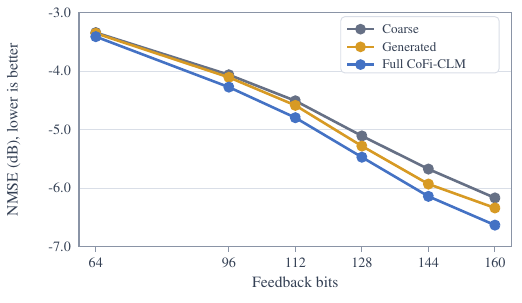}
    \caption{Rate-wise seen-scenario NMSE of the direct coarse branch,
    CLM-generated branch, and full CoFi-CLM. Lower values indicate better
    reconstruction.}
    \label{fig:paths}
\end{figure}

\subsection{CLM Generation and Path Complementarity}
Fig.~\ref{fig:paths} separates the reconstruction directly supported by the
reported coarse-token indices from the CLM-generated reconstruction and the complete
CoFi-CLM output on the seen-scenario test set. At 128 bits, the generated
branch improves the coarse-branch NMSE from $-5.1080$ to $-5.2810$ dB. This
0.173 dB improvement shows that the CLM supplies fine-scale information beyond
direct coarse decoding. Learned fusion further reduces the NMSE to $-5.4708$
dB, yielding an additional 0.190 dB improvement over the generated branch. At
160 bits, the two standalone branches differ by approximately 0.171 dB,
yet the full model improves the coarse branch by 0.465 dB. Thus, CLM
generation provides useful complementary information for coarse reconstruction.

\begin{table}[t]
    \centering
    \caption{128-Bit Deployment Complexity per CSI Sample}
    \label{tab:complexity}
    \footnotesize
    \setlength{\tabcolsep}{2pt}
    \begin{tabular*}{\columnwidth}{@{\extracolsep{\fill}}lrrrrrr@{}}
        \toprule
        \multirow{2}{*}{Model} & \multicolumn{2}{c}{UE}
        & \multicolumn{2}{c}{BS} & \multicolumn{2}{c}{BS share (\%)}\\
        \cmidrule(lr){2-3}\cmidrule(lr){4-5}\cmidrule(l){6-7}
        & Params. & MACs & Params. & MACs & Params. & MACs\\
        \midrule
        Small CLM & 0.811M & 0.224G & 4.302M & 0.828G & 84.1 & 78.7\\
        Large CLM & 0.811M & 0.224G & 30.071M & 2.355G & 97.4 & 91.3\\
        \bottomrule
    \end{tabular*}
\end{table}

\subsection{BS-Centric Complexity and Deployment}
CoFi-CLM places the Transformer, both decoders, and fusion module at the BS,
leaving coarse tokenization and index transmission at the UE.
Table~\ref{tab:complexity} quantifies this deployment through theoretical
online complexity per $2\times32\times32$ CSI sample with batch size one.
One multiply--accumulate operation (MAC) counts as two floating-point
operations. The offline fine encoder is excluded, and the shared coarse
codebook is counted at both ends. The results show that the BS accounts for
most parameters and MACs. Increasing CLM capacity therefore raises BS-side
cost while leaving UE computation and storage unchanged.

\section{Conclusion}
This paper formulated finite-bit CSI recovery as predicting distributions over
fine-token indices conditioned on coarse-token indices and proposed CoFi-CLM.
The UE reports indices from a learned coarse VQ codebook, while the BS-side CLM
predicts distributions over the fine codebook and fuses generated and direct
reconstructions. Across 64--160 feedback bits,
CoFi-CLM outperforms the compared methods at every evaluated rate under the
considered protocol. Results on unseen scenarios support the cross-scenario
generalization of CoFi-CLM under the considered channel model. The
parameter-matched comparison indicates that the performance gain cannot be
explained by model size alone. The reconstruction-path ablation demonstrates
the complementarity between CLM generation and coarse decoding. Moreover,
scaling the BS-side CLM further improves reconstruction accuracy without
increasing UE-side computation or storage.

Future work will evaluate the framework under broader channel conditions and
investigate rate-adaptive models with lower BS inference cost.

\bibliography{Reference}

\end{document}